\documentclass[preprint,aps,prd,nofootinbib,12pt]{revtex4}
\usepackage{amsmath}
\usepackage{graphicx}
\begin{document}

\title{Falling chains as variable mass systems: theoretical model and experimental analysis}
\author{C\'elia A. de Sousa, Paulo M. Gordo and Pedro Costa}
\affiliation{Department of Physics, University of Coimbra, P-3004-516 Coimbra, Portugal}
\email{celia@teor.fis.uc.pt}

\begin{abstract}

 In the present paper we revisit,  theoretical and experimentally, the fall of a
folded U-chain and of a pile-chain. The model calculation implies the division of the
whole system into two subsystems of variable mass, allowing to explore the role of
tensional contact forces at the boundary of the subsystems. This justifies, for instance,
that the folded U-chain falls faster than the acceleration due to the gravitational
force. This result, which matches quite well with the experimental data independently of
the type of chain, implies that the falling chain is well described  by energy
conservation. We verify that these conclusions are not observed for the pile-chain
motion.
\end{abstract}

\maketitle

\section{Introduction}

The dynamics of variable mass systems has been studied by professors and  researches
since the 19th century \cite{Cayley}, becoming a particular branch within classical
mechanics. From now on, applications involving variable mass systems are distributed over
a wide range of different areas of knowledge, such as in rocket theory \cite{Meirovitch},
astronomy \cite{Kayuk}, biology \cite{Canessa2007}, econophysics \cite{Canessa2009},
robotics \cite{Djerassi}, mechanical and electrical machinery \cite{Cveticanin2010} and
engineering in general.
However, despite these successful applications, even today one can find, in the
specialized and/or pedagogical  literature,  apparent paradoxes that allow for
discussions on the fundamentals of the variable mass system dynamics.
 In practice, we verify that the study of these systems is a challenge for students, and enhances their
understanding of  laws of dynamics for systems of particles.
 These are good
arguments to the  introduction of this theme  to  students in the scientific areas. To
this purpose,  ropes or chain systems  have been considered as didactic examples, being
studied  from different points of view. In this context, the published suggestions are
carried out by using several  procedures, such as the concept of momentum flux
\cite{Siegel,Sousa2002},
 the generalization of  Newton's second law \cite{Sousa2004},  analytical
 methods \cite{Wong}, and numerical simulations \cite{Tomaszewski}.
  Some of these methodologies are adequate frameworks for beginning students of
  physics \cite{Ped}.

 In the present paper, we consider two well known examples of such systems: a folded
chain initially suspended from a rigid support by the two ends placed closed together,
and then one of the ends is released; and a chain, part of which is hanging off the edge
of a smooth table.
We refer briefly to these systems as U-chain and pile-chain \cite{Ruina}.
The example of the pile-chain is equivalent to a pile of  chain falling through  a smooth
hole. However, we prefer to consider here the pile-chain falling off from the edge of a
support because it is  easier to solve technical difficulties in the experimental
procedure.

 Recent  experimental observations that the free end of
 the U-chain falls faster than $g$ \cite{Tomaszewski,Hamm,Calkin} have contributed   to an
increasing interest on the behavior of systems such as ropes and chains. Such a behavior
admits that the falling chain is a conservative system. For  completeness and comparative
purposes, we also include the model calculation where the energy conserving assumption is
not considered {\em a priori}. In this case the acceleration of the falling arm  of
 the U-chain is $g$.

 In the pile-chain configuration, the hypothesis of conservation of mechanical energy  leads to the  value
$g/2$ for the acceleration of the chain tip \cite{Sousa2004}. If the energy conserving
scenario  is not assumed {\em a priori},  the value  $g/3$  is obtained \cite{Cayley}.

As already referred by other authors \cite{Wong}, there are significant differences
between the behavior of chain systems in  the two configurations. Whereas in the falling
U-chain it is possible to obtain a link-by-link mass transfer at the  fold of the chain,
such a behavior is difficult to realize in a real chain falling from a resting heap
through a hole.  In fact,  it is very difficult to get a well arranged pile in order to
allow for a steady motion. So, it is tricky to confirm experimentally which model is more
reliable to describe such systems in this configuration.
 Some authors also refer to an additional complexity which consists in the lift  above the
 platform before  the fall of the chain, a feature  justified by the energy conserving  assumption as
  will be discussed later.

 Newton's second law for variable mass systems have been successfully  applied to chains and ropes,
 independently of the model assumption, i.e., by either considering or not the energy
 conservation approach
 \cite{Sousa2004}.
 The method, here included for the sake of completeness,    considers the whole system divided into
 two subsystems of variable mass. The motion is  one-dimensional, and the forces at
 the boundary of the two subsystems, one of which is at rest, can be calculated. This
 explains, for instance, why the acceleration of the falling chain is larger than $g$  in
 the  U-chain  when the energy conserving approach is assumed.

 A special attention is dedicated to
the pile-chain configuration. As already referred, some aspects of this problem
 have been solved in the literature by two distinct
approaches. Our main purpose is to discuss the  { two solutions $a=g/3$ and $a=g/2$,
trying to confront these theoretical  results with the experimental data. Bearing in mind
the important role played by   tensional contact forces, we also show that the analysis
of the variable mass system, as a redistribution of mass between the two subsystems, is
reliable to study that tension.
We remind  that some treatments analyse the problem using energy conservation only,
without deriving  equations of motion.

We include in section 2 a summary of the methodology used  to study variable mass
systems, and in section  3 we present the experimental setup.  Two illustrative examples
are studied theoretical and experimentally  in sections 4 and 5.
Section  6 contains the concluding remarks.

\section {Theoretical background: Newton's second law for variable mass systems}

 The equations of motion of the falling chains can be obtained by considering a closed
 one-dimensional
 system of constant total mass composed by two open subsystems which exchange mass. So, the systems  considered in this paper have  the
characteristics indicated below.

\begin{itemize}
   \item [(i)] The whole system, a constant mass one, can be divided into two (variable
   mass) subsystems I and II with masses $m_{{\rm I}}$ and $m_{{\rm II}}$, respectively.
    To the whole system the traditional  Newton's second
    law, $F=(m_{\rm I}\,+\,m_{{\rm II}})\,a_{\rm cm}$, where $a_{\rm cm}$ is the acceleration
    of the center of mass, or $F\,=\,{{\rm d}p}/{{\rm d}\,t}$, in terms of linear momentum, are valid.
   \item [(ii)] One of the subsystems (I) is  moving with velocity $v$, and the other (II) is  at rest.
   \item [(iii)] The velocity of the mass being transferred between the subsystems is
$u=v$.
 \end{itemize}

Under these assumptions two expressions  can be obtained for each subsystem (see
\cite{Sousa2004} for details)

\begin{equation}
{ F}^{({\rm I})}\,=\,\frac{{\rm d}}{{\rm d}\,t} \,(m_{\rm I}\,u)\,-\,u\,\frac{{\rm d}\,m_{\rm I}}{{\rm
d}\,t},\label{var1}
\end{equation}
and

\begin{equation}
{ F}^{({\rm II})}\,=\,\frac{{\rm d}}{{\rm d}\,t} \,(m_{{\rm II}}\,v)\,-\,u\,\frac{{\rm
d}\,m_{{\rm II}}}{{\rm d}\,t},\label{var2}
\end{equation}
where ${ u}\,{\rm d}m_{{\rm II}}/{{\rm d} t}\,({ u}\,{{\rm d}m_{\rm I}/{{\rm d} t}})$  is
the rate at which momentum is carried into (away)  the system of mass $m_{\rm
II}\,(m_{\rm I})$ {---} ``momentum flux''.

As both equations (\ref{var1}) and (\ref{var2}) have a common generic structure, hereafter the symbols
I and II are dropped allowing to write
 Newton's second law for variable mass
systems  in the form

\begin{equation}
\frac{{\rm d}\,{ p}}{{\rm d}\,t}\,=\,{ F}\,+\,{ u}\,\frac{{\rm d}\,m}{{\rm
d}\,t},\label{var}
\end{equation}
where  $m$ is the  instantaneous mass, ${ p}\,=\,m\,{v}$
 its linear momentum, and
  ${ F}$   is the net external force acting upon the  variable mass system.

 \section {Experimental setup}

The  experimental measurements were performed using three chains: one ``ball chain'' and
two sizes, medium and large, of normal ``linked chains''. The ball chain consists of
stainless-steel identical segments made from rods and spheres attached to each other. The
two loop chains are made of torus-shaped links.  The dimensions of the torus-shaped links
of the medium (large) linked chain are: length 11 mm (20 mm)  and width 18 mm (24 mm).
Other characteristics of the chains are given in table~\ref{tab:I}.

\begin{table}[h]
\begin{center}
\begin{tabular}{|c|c|c|c|c|}
  \hline\hline
  Chain & \hskip0.3cm l(m) \hskip0.3cm & \hskip0.3cm m(kg) \hskip0.3cm & \hskip0.3cm $\lambda$(kg/m)\hskip0.3cm
   &Number of links \\
  \hline
  ball chain & 1.830(1) & 0.0395(1) & 0.0216(1) &419 \\
  medium loop chain & 1.859(1) & 0.1860(1) & 0.1001(1)  &115\\
  large loop chain  & 1.738(1) & 0.4029(1) & 0.2318(1)  &99\\
  \hline
  \hline
\end{tabular}
\caption{\label{tab:I} Characteristics of chains used in the experiments.}
\end{center}
\end{table}

The experimental setup is based on the force sensor from Pasco model CI-6537  (maximum
force of 50 N), that is usually available in  most  educational laboratories. The signal
from the force sensor is monitored using a ScienceWorkshop interface, model 750 also from
Pasco, connected to a personal computer, that makes use of DataStudio (version 1.9.7r12)
software to control the data acquisition. The typical time duration of the events that we
are investigating is of the order of 1 second or less. The rate acquisition signals of 1
kHz or higher is sufficient to reveal the details of the  dynamics during the experiment.
We acquired the signals at 1, 2, 4 and 10 kHz,  which gave similar information
 in all cases except in the number of  data points to manipulate.
We  have also used a photogate head, model ME-9204B from Pasco, to define  the important
and critical instant $t=0$.

For the U-chain experiment, one of the ends of the chain was attached to the hook of the
fixed force sensor and the other end was released from the same height. The system is in
vertical position and the sensor measures the net force by the chain on sensor's hook.

For the pile-chain experiment two different devices were used: (A) a flat polished
squared wood table with  21.5 cm of  length, fixed to the force sensor, from which the
pile-chain falls down; and (B) a ceramic ashtray of diameter 8.58 cm, with lateral
U-shaped overture of width approximately 2.5 cm, also fixed to the force sensor. In both
cases, a special accessory was used to attach the devices to the force sensor.  The
beginning of the chain's fall was monitored by the photogate sensor which was placed near
the border of the table (overture of the ashtray).

\section{Analysis of the U-chain}

 A folded uniform chain with length $l$ and mass  per unity of length $\lambda$ is
considered.
 In figure~\ref{fig:U-chain}  we illustrate the configuration of the system at
$t=0$ and $t\neq 0$, where  $x$ is a generic position of point B. As illustrated, the
axis $x$ points downwards and, for convenience, the system is divided into two subsystems
I and II (see figure 1 and table~\ref{tab:II}).

\begin{figure}[h]
\begin{center}
\scalebox{0.55}{\includegraphics{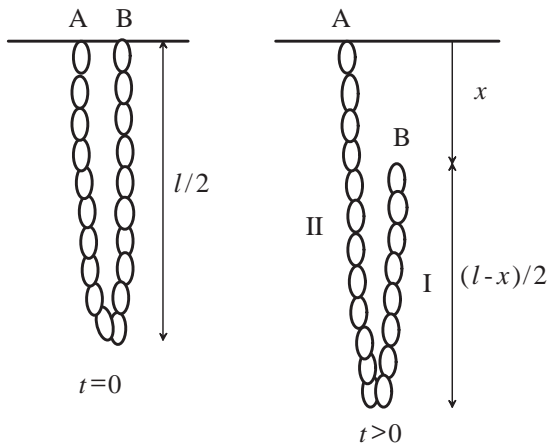}} \caption{\label{fig:U-chain} Scheme of the
U-chain configuration.}
\end{center}
\end{figure}

\subsection {Energy conserving assumption}

 We start by considering   that the mechanical energy of the U-chain is conserved
\cite{Wong,Tomaszewski,Hamm}.
 Within this model
assumption, the acceleration of the falling chain is not known, neither are the  tension
forces indicated in table~\ref{tab:II}. Consequently, it is not convenient to start the
resolution of the system by using  Newton's second law for constant or variable mass
systems.

\begin{table}[h]
\begin{center}
\begin{tabular}{|c|c|c|c|c|c|}
\hline
System& $m$&$p$&$F$&$u$& $u{\rm d} m/{\rm d}t$\\
\hline
 I&$\frac{\lambda}{2}\,(l\,-\,x)$ &$\frac{\lambda}{2}\,(l\,-\,x)\,v$&$\frac{\lambda}{2}\,(l\,-\,x)\,g\,+\,T_1(x)$&$v$&$- \frac{\lambda}{2}\,v^2$\\
\hline
II &$\frac{\lambda}{2}\,(l\,+\,x)$ &$0$& $\frac{\lambda}{2}\,(l\,+\,x)\,g\,+\,T\,(x)\,+\,T_2 (x)$&$v$&$ \frac{\lambda}{2}\,v^2$\\
\hline
I+II &$\lambda\,l$ &$\frac{\lambda}{2}\,(l\,-\,x)\,v$& $\lambda\,l\,g+\,T\,(x)$&{---} &{---}\\
\hline
\end{tabular}
\caption{\label{tab:II} Summary of the relevant physical quantities for the U-chain. See
the text for notation.}
\end{center}
\end{table}

 The variation of
kinetic and potential energies are easily calculated and, from the conservation of energy
$\Delta E=0$ we obtain \cite{Marion}

\begin{equation}
\frac{\lambda}{4}\,(l\,-\,x)\,v^2\,-\,\frac{\lambda}{4}\,g\,x\,(2\,l\,-\,x)\,=\,0.
\end{equation}
Eliminating $v^2 (x)$ from the last equation, we find
\begin{equation}
v^2\,(x)\,=\,g\,x\,\frac{2\,l\,-\,x}{l\,-\,x}.\label{v2}
\end{equation}

 The acceleration of the falling arm of the chain   can be obtained  directly from this equation by applying the
  following mathematical procedure:

\begin{equation}
a\,=\,\frac{{\rm d}\,v}{{\rm d}\,t}\,=\,\frac{{\rm d}\,v}{{\rm d}\,x}\,\frac{{\rm
d}\,x}{{\rm d}\,t}\,=\,\frac{1}{2}\,\frac{{\rm
d}\,v^2}{{\rm d}\,x},\label{id}
\end{equation}
 giving

\begin{equation}
a\,=\,g\,+\,\frac{g}{2}\,\,\frac{x\,(2\,l\,-\,x)}{(l\,-\,x)^2}\,=\,g\,+\,\frac{v^2}{2\,(l-x)}.\label{acx}
\end{equation}

The application of  Newton's second law to the whole system, for instance in the form
$F={{\rm d}\,p}/{{\rm d}\,t}$, allows for the tension, $T\,(x)$,  acting on the chain by
the support (see table~\ref{tab:II}). In fact, we can write

\begin{equation}
\frac{{\rm d}}{{\rm
d}\,t}\,\left(\frac{\lambda}{2}\,(l\,-\,x)\,v\,\right)\,=\,\lambda\,l\,g\,+\,T
(x).\label{aa}
\end{equation}
We substitute  (\ref{v2}) and (\ref{acx}) into (\ref{aa}) and find

\begin{equation}
T\,(x)\,=\,-\,\frac{\lambda}{2}\,\,g\,(l\,+\,x)\,-\,\frac{\lambda}{4}\,\,g\,x\,\frac{2\,l\,-\,x}{l\,-\,x},\label{ten}
\end{equation}
where the minus sign indicates that  this force acts upward, as it must be.

It is instructive to rewrite this equation  in the form

\begin{equation}
T\,(x)\,=\,-\,\frac{\lambda}{2}\,g\,(l\,+\,x)\,-\,\frac{\lambda}{4}\,v^2,
\end{equation}
 showing that this tension force results from the instantaneous ``weight'' of the chain,
 i.e., from the weight of the static side of the
 chain (II) at a given instant, and an additional ``dynamic weight'' which depends on the velocity of
 the falling side (I).

The contact at the bottom provides a tension force $T_1(x)$ that can be obtained by
applying  Newton's second law (\ref{var}) for subsystem I. In fact, for this subsystem
one has (see table~\ref{tab:II} and figure~\ref{fig:U-chain})

\begin{equation}
m\,=\,\frac{\lambda}{2}\,(l\,-\,x), \hskip0.5cm
p\,=\,\frac{\lambda}{2}\,(l\,-\,x)\,v,\hskip0.4cm {\rm and} \hskip0.4cm u\,\frac{{\rm d}
m}{{\rm d} t}\,=\,-\,\frac{\lambda}{2}\,v^2,
\end{equation}
allowing for

\begin{equation}
T_1\,(x)\,=\,\frac{\lambda}{4}\,g\,x\,\frac{2\,l\,-\,x}{l\,-\,x}.\label{t1}
\end{equation}

 We point out that this expression for $T_1(x)$ can be rewritten in the form
 $T_1\,(x)\,=\,\lambda\,v^2\,/4$,   result that can also be obtained using the Lagrangian
 formalism \cite{Wong}.

The downward tension $T_1(x)$ in the chain on the falling side pulls  the chain down in
addition  to the  gravitational  force. So, with this methodology, the tension at the
fold comes naturally.

 The same procedure applied to subsystem II allows for the tension
$T_2\,(x)\,=\,-\,T_1\,(x)$, satisfying    Newton's third   law.

To compare the model calculations with the experimental results we integrate
(\ref{v2}). To avoid the
 complexity, which comes from the solution of $x(t)$ in terms of elliptic integrals, we  compute this
 function of time numerically. With this procedure we obtain $x\,=\,x(t)$ that, together  with  (\ref{ten}),
  allows to attain $T\,=\,T (t)$.

\subsection {Free fall  assumption}

Now,  the energy conserving assumption is not considered {\em a priori}. Instead, we
assume that  the falling part of the chain (subsystem I) falls with an acceleration $g$,
as in free fall \cite{Marion}.
 This model assumption  presented in \cite{Sousa2004} with other geometry  is  briefly presented here by completeness
  and comparative purposes.

So, by applying (\ref{var}) to subsystem I (see table 2 and figure 1) we conclude that $T_1\,(x)\,=\,0$.

An analogous procedure applied to subsystem II, together with the condition
$T_2\,(x)\,=\,-\,T_1\,(x)\,=\,0$,  allows to obtain  the tension at the support

\begin{equation}
T\,(x)\,=\,-\,\frac{\lambda}{2}\,g\,(l\,+\,x)\,-\frac{\lambda}{2}\,v^2.\label{tt1}
\end{equation}

The expression of the squared velocity as a function of $x$ can  be calculated by using equation (\ref{id}) with $a=g$, allowing for

\begin{equation}
v^2(x)\,=\,2\,g\,x.
\end{equation}

Finally, from the two last equations  we obtain
\begin{equation}
T\,(x)\,=\,-\,\frac{\lambda}{2}\,g\,l\,\left (1\,+\,3\,\frac{x}{l}\right).\label{tt1}
\end{equation}

To compare this model calculation   with the experimental results the well known solution
$x (t) = 1/2\,g\,t^2$ is used to obtain the tension

\begin{equation}
T\,(t)\,=\,-\,\frac{\lambda}{2}\,g\,l\,\left
(1\,+\,\frac{3}{2}\,\,\frac{g}{l}\,t^2\right).\label{ttt}
\end{equation}

\subsection{Comparison with the experimental results}

The theoretical calculations obtained in the previous subsections are going to be
compared with the experimental results. We start releasing the ball type chain. In
figure~\ref{fig:U-signal} we present  the typical signal from the  force sensor,
measuring the force by the falling chain in the  sensor. Similar curves were observed for
the other two loop chains, except in some details such as   the magnitude of the measured
force due to  different  weight of the chains.  The represented values of the force are
negative because of the convention signal of the sensor: it is positive (negative) when
there is a compression (distension) of the sensor.

\begin{figure}[h]
\begin{center}
\scalebox{0.50}{\includegraphics{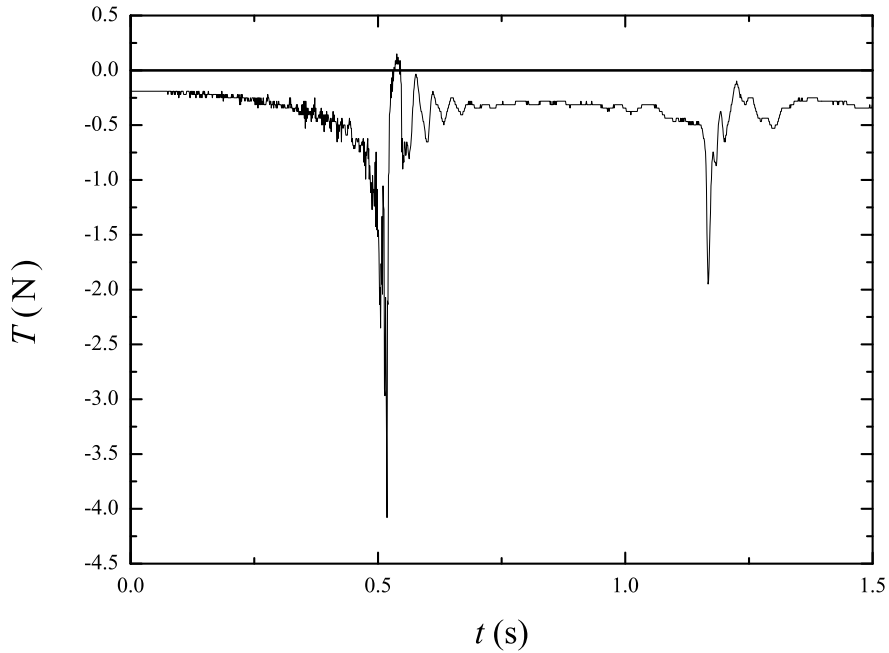}} \caption{\label{fig:U-signal} Tension at
the fixed end of the chain   as a function of time for the ball chain.}
\end{center}
\end{figure}

For $t=0$, the value observed in the sensor corresponds to  half of the weight of the
ball chain, $T_0=- 0.193$ N,   and the fall of the chain occurs in approximately $0.5$~s,
when maximum value for the tension force  is achieved. The behavior of the tension $T
(t)$ for  $t>0.5$ s  reflects the ricochet effect of the chain in the sensor. The maximum
value of the tension  achieved at the end of the fall is  larger than 50 times the
chain's weight. This agrees with the energy  conserving assumption for this system, which
predicts a very high velocity (see (\ref{v2})) of the chain at the end of the fall and,
consequently, a very strong variation of $T$  at the end of the movement. For instance,
$T/T_0\,=\,51.9$ at $x/l\,=\,0.99$  in this model assumption, whereas in the free fall
assumption $T/T_0\,=\,4$ at $x=l$ (see  (\ref{tt1})).
We remember that   the force sensor  reads the instantaneous ``weight'' of the chain,
which  is larger than the weight of the chain actually at rest at a given instant.

\begin{figure}[h]
\begin{center}
\scalebox{0.60}{\includegraphics{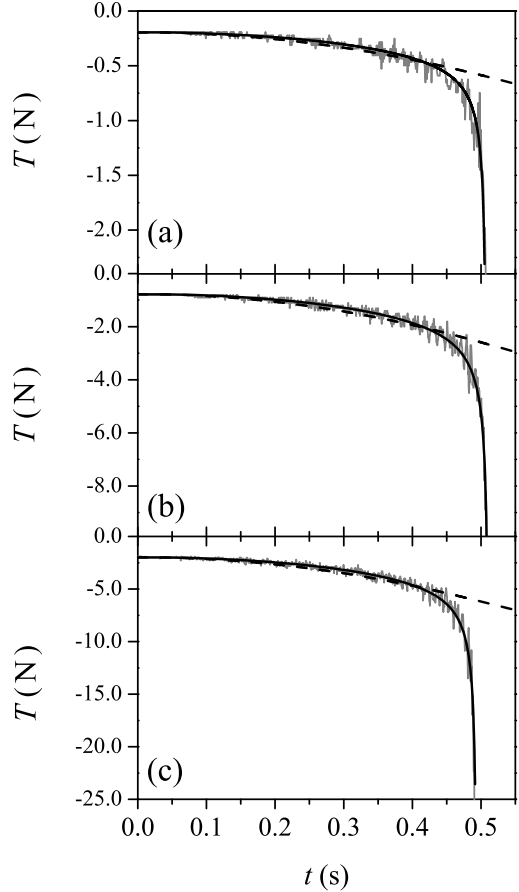}} \caption{\label{fig:U-graph} Tension at
the fixed end of the chain as a function of time for: (a) ball chain, (b) medium loop
chain, and (c) large loop chain. The solid (dashed) curve represents the theoretical
result obtained by energy conserving (free fall) assumption.}
\end{center}
\end{figure}

In figure~\ref{fig:U-graph},  the theoretical results   of the tension $T (t)$ as a
function of $t$  are plotted during the fall of the chain together with the experimental
results, for the cases: (a) ball chain,  (b) medium loop chain, and  (c) large loop
chain.
  As referred in   subsection 4.1,   the theoretical curves in the energy conserving approach
have been obtained numerically. The observed fluctuations in the value of $T$  are mainly
due to the discrete shape of the chains especially in the case of loop chains as can be
seen in figure~\ref{fig:U-graph}. A larger loop chain corresponds to higher amplitude of
these fluctuations (notice the difference of scale in the three cases).
It can also be
seen that the  fall time for all chains are quite identical, because of the similarity of
the length of the chains (see table~\ref{tab:I});  the shorter time is observed for the
large loop chain, the one with the shortest   length (about minus 10 cm).
For the large and medium loop chains, the maximum values of $T$ achieved   (not
represented in the figure) are several orders of magnitude higher than  their weight. The
correct value could not be determined precisely due to the limit of $50$ N for the
maximum force of the sensor.

 It can be seen that the theoretical result given by the solid curve  and the
experimental data fit quite well, and we can conclude that the dynamics of the falling
chain is very well described by the conservation of energy assumption, independently of
the type of chain.
 This is explained by the continuous interference (no broken contacts) at the fold of
the chains during the motion. Consequently,  the acceleration of the moving part of the
chain is  higher than $g$ at the end of the movement.  The falling chain is divided into
two subsystems: the almost motionless part attached to the support and the moving part.
The falling chain lowers its potential energy on the account of the kinetic energy of the
continuously shorter part of the chain. Because the mass of the latter decreases, its
velocity grows significantly.
We remember that,   as it was reported early \cite{Calkin}, if the   initial distance
between the ends of the chain is  moderated, the contribution to the kinetic energy from
the horizontal motion is a small fraction of that due to the vertical motion, and the
system can appropriately be treated in the limiting one-dimensional motion.
We can conclude that real chains behave like a flexible and  inextensible  conservative system.

\section{Analysis of the pile-chain}

Now we consider the fall of  a pile-chain from the edge of a platform. One end of the
chain falls and pulls the chain after it in a steady motion (see
figure~\ref{fig:pile-chain}). The chain starts moving from rest at $x=0$ and the $x$ axis
points downward. This geometry is equivalent to the fall of a pile of chain placed just
above a hole in a platform. In fact, both problems reflect the same physical reality,
however, the first case is more easily implemented by normal pieces like a flat platform
or an ashtray.
As already referred, some aspects  of this  problem  have been solved in the literature
by considering two scenarios. Our purpose is to discuss both solutions for the force on
the platform, trying to confront the results with experimental data.

\begin{figure}[h]
\begin{center}
\scalebox{0.65}{\includegraphics{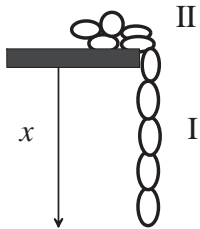}} \caption{\label{fig:pile-chain} Scheme
of the  falling pile-chain.}
\end{center}
\end{figure}

\subsection {Energy conserving assumption }

We consider that the chain is a one-dimensional system and we start with  the assumption
of  conservation of mechanical energy.
The acceleration of the falling chain is
not known, as well as the force $N(x)$ acting on the chain  by the platform.

Referring to  figure~\ref{fig:pile-chain}, we see that  from the conservation of energy
we obtain

\begin{equation}
\frac{1}{2}\,\lambda\,x\,v^2\,-\,\frac{1}{2}\,\lambda\,g\,x^2\,=\,0,
\end{equation}
which gives the square of the velocity $v^2$ in
terms of  $x$:
\begin{equation}
v^2(x)\,=\,g\,x.\label{v22}
\end{equation}

The acceleration of the falling rope can be obtained:

\begin{equation}
a\,=\,\frac{{\rm d}v}{{\rm d}t}\,=\,\frac{1}{2}\,\frac{{\rm d}\,v^2}{{\rm
d}\,x}\,=\,\frac{g}{2}.
\end{equation}

\begin{table}
\begin{center}
\begin{tabular}{|c|c|c|c|}
\hline
System& $m$&$p$&$F$\\
\hline
 I&${\lambda}\,x$ &$\lambda\,x\,v$&$\lambda\,x\,g\,+\,T_1(x)$\\
\hline
II &$\lambda\,(l\,-\,x)$ &$0$& ${\lambda}\,(l\,-\,x)\,g\,+\,T_2\,(x)\,+\,N\, (x)$\\
\hline
I+II &$\lambda\,l$ &${\lambda}\,x\,v$& $\lambda\,l\,g+\,N\,(x)$\\
\hline
\end{tabular}
\caption{\label{tab:III} Summary of the relevant physical quantities for the pile-chain.
See the text for notation.}
\end{center}
\end{table}

Now, we apply  Newton's second law, for instance in the form $F={\rm d}p/{\rm d}t$,  to
the whole system of constant mass $\lambda\,l$.

The net external forces  acting on this system  are the downward gravity and the normal
force by the platform, $N(x)$, allowing for the equation of motion (see
table~\ref{tab:III})
\begin{equation}
\frac{{\rm d }}{{\rm d} t}\,(\lambda\,x\,v)\,=\,\lambda\,l\,g\,+\,N\,(x).
\end{equation}

Straightforward calculation, which includes the substitution of $v$ and $a$ in the
previous equation, provides the  total  force by the platform  on the chain

\begin{equation}
N\,(x)\,=\,-\,\lambda\,\left(l\,-\,\frac{3}{2}\,x\right)\,g,\label{nxc}
\end{equation}
where the minus sign indicates that this force acts upward, as it must be. This equation
also  shows that $N(x)=0$ when $x=2l/3$, indicating the domain of validity of the
previous equation in the interval $0\leq x\leq 2 l/3$.

As in the previous example, the application of  Newton's second law to the variable mass
system I (II) allows the calculation of the tension $T_1(x)$ ($T_2(x)$) at the boundary
of subsystem I (II). In fact, for  subsystem I one has (see table~\ref{tab:III} and
figure~\ref{fig:pile-chain})

\begin{equation}
m\,=\,{\lambda}\,x, \hskip0.5cm p\,=\,{\lambda}\,x\,v,\hskip0.4cm {\rm and} \hskip0.4cm
u\,\frac{{\rm d} m}{{\rm d} t}\,=\,\lambda\,v^2.\label{mpu}
\end{equation}

  Using (\ref{mpu}) in (\ref{var})  with $F$ given from table 3, leads to

 \begin{equation}
 T_1\,(x)\,=\,-\,\lambda\,x\,g/2\,=\,-\,\lambda\,v^2/2.
 \end{equation}
 With this procedure the tension at the boundary of the subsystems
comes naturally.

To compare with the experimental results, we calculate  the expression for $N(t)$ by
inserting  the equation of motion $x\,=\,\frac{1}{4}\,g\,t^2$ into   (\ref{nxc}), and
thus
\begin{equation}
N\,(t)\,=\,-\,\lambda\,l\,g\,\left(1\,-\,\frac{3}{8}\frac{g}{l}\,t^2\right).\label{ntc}
\end{equation}

We notice that the problem at hand has four unknowns: the acceleration $a$, the normal
$N(x)$, and the tension  forces  $T_1(x)$ and $T_2(x)$ acting through the boundary of the
subsystems. As $T_1(x)$ and $T_2(x)$ satisfy  Newton's third law, we have, effectively,
three unknowns to be determined by three equations: the law of conservation of energy,
and for instance,  Newton's second law applied to the whole system and   Newton's second
law  for variable mass system I or II.

\subsection {Tait-Steele solution}

 Now, we consider  that mechanical energy is not conserved {\em a priori}. As the number of unknowns holds, the
system is undetermined.  To make the equations determined while preserving their
simplicity, some conditions must be imposed.

Tait and Steele solve the problem by using Newton's second law in the form \cite{Tait}

\begin{equation}
\frac{{\rm d}}{{\rm d} t}\,(\lambda\,x\,v)\,=\,\lambda\,x\,g.\label{cay}
\end{equation}
This equation can be written as a first-order linear equation in $v^2$
\begin{equation}
v\,x\,\frac{{\rm d} v}{{\rm d} x}\,+\,v^2\,=\,x\,g.
\end{equation}
The solution of this equation is

\begin{equation}
v (x)\,=\,\left(\frac{2}{3}\,g\,x\right)^{1/2},\label{vv}
\end{equation}
allowing for the well known result $a=g/3$.

However,  the falling chain is a variable mass system for which  Newton's second law
$F={\rm d} p/{\rm d}t$ is not {\em a priori} valid, unless the following conditions are
imposed

\begin{equation}
 u=0, \hskip0.5cm {\rm and}\hskip0.5cm T_1(x)=0 ,\label{cond}
\end{equation}
 i.e., particles enter/leave the variable
mass system with zero velocity, and there is no interference between adjacent links in
the boundary of both subsystems.

Alternatively, it is also possible to obtain  (\ref{cay}) by applying Newton's second
law  to subsystem I (see (\ref{var}) and  table~\ref{tab:III}), with the more
realistic condition $u=v$, if we also impose an {\em empirical}  condition on $T_1$,
i.e.,

\begin{equation}
 u=v, \hskip0.5cm {\rm and}\hskip0.5cm T_1(x)=-\lambda\,v^2.\label{cond2}
\end{equation}

With these conditions, the effect of the upward tension $T_1$ is canceled by the
``momentum flux'', $u\,{\rm d} \,m/\,{\rm d}\,t$,  carried away from  subsystem I.

In reality, we must consider $u=v$, and the tension $T_1(x)$ is undetermined.
Consequently,  Newton's second law  (see (\ref{var})) applied to subsystem I allows
 for
\begin{equation}
a \,=\,g\,\,+\,\frac{T_1\,(x)}{\lambda\,x}.\label{ac2}
\end{equation}

 This problem is
 absent in the U-chain description when the conservation of energy is not considered {\em a priori},
 because the condition $a=g$  for the acceleration of the falling
 chain is assumed. This provides the
 extra equation necessary to solve the problem, and   $T_1(x)=0$ comes naturally.

Finally, to test the  reliability of the solution $a=g/3$, we are going to obtain the
force on the chain by the platform, $N(t)$, to be compared with the experimental data.

To this purpose, applying   Newton's second law to the whole constant mass system, we
write

\begin{equation}
\frac{{\rm d}}{{\rm d} t}\,(\lambda\,x\,v)\,=\,\lambda\,l\,g\,+\,N(x).\label{vmm}
\end{equation}

Comparing  (\ref{vmm}) and (\ref{cay}), we find now the   expression

\begin{equation}
N (x)\,=\,-\, \lambda\,g\,(l\,-\,x),\label{norx}
\end{equation}
which can be expressed in terms  of time:

\begin{equation}
N\,(t)\,=\,-\,\lambda\,l\,g\,\left(1\,-\,\frac{1}{6}\frac{g}{l}\,t^2\right).\label{nort}
\end{equation}

\subsection{Comparison with the experimental results}

To perform this experiment we have elaborated two configurations: (A) a flat polished
squared wood table and (B) a ceramic ashtray with lateral U-shaped overture from which
the chain falls down.

The  instant $t=0$ has been  determined by a photogate sensor, which was placed near the
border of the table (overture of the ashtray). This configuration enables us to identify
the critical instant of the beginning of the chains fall  which is difficult to check
otherwise.

In figure~\ref{fig:ball-graph}, we   show the behavior of the normal $N$   for the ball chain in the
experimental configuration setup (A).  The theoretical results  obtained by  the energy
conserving approach (\ref{ntc}) and  by the Tait-Steele solution  (\ref{nort}),
represented by the solid and dashed curves, respectively, are also shown.

\begin{figure}[h]
\begin{center}
\scalebox{0.50}{\includegraphics{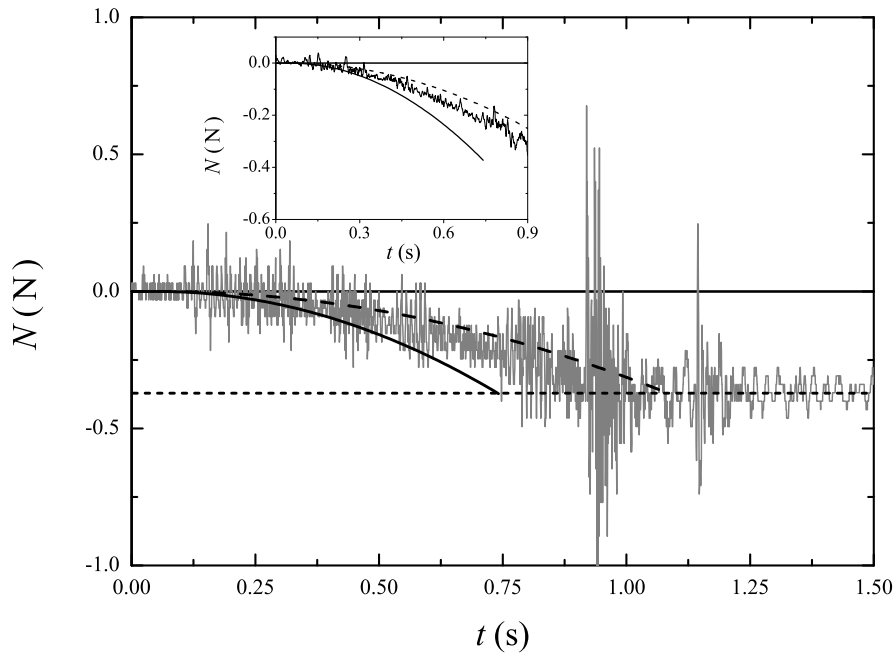}} \caption{\label{fig:ball-graph} Normal
force as a function of time for the ball chain on the platform. The theoretical results
for energy conserving approach (solid curve) and for Tait-Steele solution (dashed curve)
 are also included. Inset: Filtered signal. The signal is smoothed by averaging over an
interval of 10 ms.}
\end{center}
\end{figure}

At  $t=0$ the value of the normal force $N$ is, of course,  the weight of the chain,
re-scaled to $N=0$.  Consequently, after the fall of the chain, the absolute value of $N$
represents its weight (horizontal dashed line).  The fall of the entire
chain occurs in approximately $0.9$ s.

 As can be seen in figure~\ref{fig:ball-graph}, instead of the continuous quadratic decrease in the
normal, $N(t)$, we observe relatively large fluctuations in the force signal during the
fall. These fluctuations are  a natural consequence of the discrete nature of the chain
and must be associated with the collisions  of the links with the border of the table.
This effect is clearly reinforced with both loop chains as we will see later, as a
consequence of a more pronounced discrete nature of the systems. The amplitude of the
fluctuations increases with time and the frequency of the peaks raises, as well, due to
the increase of the frequency of the collisions. In the case of the ball chain, this
effect is less pronounced since the system is almost a ``continuous system'' as a result
of the small size and mass of the spheres that compose the chain.
The ``hidden'' ideal force $N$ can then be revealed by using a numerical filtering
procedure, which takes the average force value during 10 ms (see
figure~\ref{fig:ball-graph}, inset).

By comparing the experimental results with the theoretical curves, we observe that the
dynamics of the ball chain is not  adequately described by any of the two models.
Particularly, the solid curve for the energy conserving assumption diverges significantly
from the experimental data. From the point of view of this energy conserving approach,
and according to  (\ref{nxc}), it was expected a normal force $ N=0$ (in the figure $N$
is equal to the weight of the chain due to the re-scale done) when $x=2\,l /3 $, which is
only observed during the experiment for $x\approx l$. In fact, just before the total fall
from the platform, the very last  small portion of the chain suffers a jump from the
table.
 When  the experiment is repeated using the  configuration setup (B),
a quite similar behavior is observed for this chain.  These features suggest that the
ball chain can provide the ``ideal'' framework to discuss the validity of the theoretical
model assumptions. As will be seen below, when the experiments are made with the loop
chains, significant differences between the data of  the two setups are observed.

The main conclusion is that the actual experimental configuration is not well described
by the  one-dimension motion of the pile-chain either assuming  energy conservation or
not and, consequently, the real chains do not act  more like a perfect flexible,
inextensible rope as it was observed in the U-chain experiments.
 In practice, it is not technically possible to accommodate the static part of the
chain in a ``single point''.   So, in order to obtain a steady motion of the falling chain,
the system must spread over a finite area, which introduces features not included in the
theoretical model assumptions.
Indeed, due to the finite size and width of the chain, the contribution to the kinetic energy from the horizontal motion of the chain
over the table is a significant fraction of that due to the vertical motion.
To treat correctly this system the full two-dimensional motion of the pile-chain has to
be considered, which is a difficult mathematical task. Furthermore, it is also impossible
to eliminate some dissipative mechanisms such as friction or interferences, and
collisions between the links of the chain.

\begin{figure}[h]
\begin{center}
\scalebox{0.55}{\includegraphics{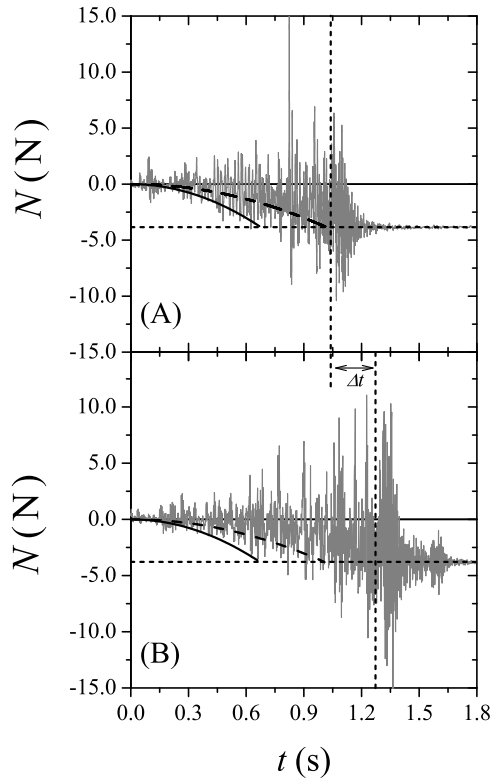}} \caption{\label{fig:loop-graph}
Comparison of the normal as a function of time for the large loop chain in the platform
(A) and in the ashtray (B). The theoretical results for energy conserving model (solid
curve) and for Tait-Steele solution (dashed curve)   are also included.}
\end{center}
\end{figure}

The dissipative mechanisms are, of course, strongly dependent on the experimental
apparatus. This is illustrated in figure 6 where  a significant difference in the results
for both configuration setups (A) and (B) for the large loop chain are shown.
Of course, in this case the experimental result is also not described by any of the two models.
As a consequence of the loop structure of this chain, the collisions between the
links are present and a concomitant lost of energy is associated with this phenomenon.
 There are also external dissipative forces like friction over the border of
the table (configuration setup (A)) or  collisions with the borders of the overture of
the ashtray (configuration setup (B)), which represents a significant loss of mechanical
energy. In the case of the setup (B), once the number of collisions is expected to be
higher, the energy lost must be also more pronounced.
This description agrees well with the difference in time, $\Delta t$, that has been
observed for the large loop chain in the configuration setups (A) and (B): the longer fall time
observed in the configuration setup (B) is associated with a higher loss of energy and,
consequently, a reduction in the speed of the chain. We notice
 that $\Delta t \approx  0.25$ s and this value represents an increase of about 25\% in the fall time.
For the medium loop chain we have observed similar results (not shown) as for the large
loop chain.

\section {Summary and conclusions}

The theoretical method suggested  allows  the discussion of falling chains as variable
mass systems. The whole system is divided into two subsystems of variable mass. With this
procedure the tension at the boundary of the subsystems comes naturally when the Newton's
second law for variable mass systems is applied.

The theoretical results provided by the energy conserving assumption have the following
characteristics:
\begin{itemize}
  \item in the falling U-chain, the contact at the bottom results in a downward tension force $T_1\,=\,\lambda\,
  v^2/4$, pulling the falling part with an acceleration larger than $g$;
  \item in the pile-chain, the contact at the top originates  an upward tension force $T_1\,=\,-\,\lambda\,v^2/2$,
   slowing down the falling part with an acceleration ($g/2$) smaller  than  $g$.
\end{itemize}

The model calculation when the energy conservation is not assumed {\em a priori}  allows
the following conclusions:
\begin{itemize}
  \item in the U-chain, it is assumed that $a=g$ and $u=v$, allowing for $T_1=0$;

  \item in  the pile-chain,  the acceleration is undetermined, unless it is assumed that
  $T_1 =0$ ($T_1=\,-\,\lambda v^2$)  and $u=0$ ($u=v$),
  which allows for $a=g/3$.
\end{itemize}

As already verified by other authors, the experimental data for the U-chain shows a good convergence with the energy conserving
approach,  independently of the  type of chain. There exists continuous interference  (no
broken contact) during the motion and, consequently, the dissipative effects of explicit
   collisions, even in a discrete chain, are negligible. In addition, one-dimensional model is a
   good approach for this configuration.

  To model the pile-chain  motion  two devices have been
used: a flat platform and a ceramic ashtray. We observe that the ball chain has a similar
behavior in both experimental apparatus, showing that it constitutes an  ``ideal'' system
to this type of analysis. The measured data with this chain shows that the system is not well described by one-dimensional motion
only.
For the loop chains  dissipative mechanisms like friction and collisions between the loop and the borders of the table and ashtray are present
and must be taken into account.

The experience with the ball chain falling from the platform is particularly important,
showing that it provides the best situation between the ideal case and the experimental
constraints, but it reveals that the system can not be modeled by one-dimensional motion
alone.

\vskip0.2cm
{\bf Acknowledgements}

 Work supported by F.C.T. and Physics Department.

\end{document}